\documentclass[aps,pre,twocolumn,groupedaddress,showpacs]{revtex4}
\usepackage{graphicx}
\usepackage{amsmath}
\usepackage{amsfonts}

\def\bF{\mathbf F}
\def\bE{\mathbf E}
\def\bF{\mathbf F}
\def\bH{\mathbf H}
\def\bJ{\mathbf J}
\def\bB{\mathbf B}
\def\bD{\mathbf D}
\def\bP{\mathbf P}
\def\bS{\mathbf S}

\def\bx{\mathbf x}

\def\bu{\mathbf u}

\def\bv{\mathbf v}
\def\bff{\mathbf f}

\def\bu{\mathbf u}

\def\bK{\mathbf K}

\def\bg{\mathbf g}
\def\b0{\mathbf 0}

\begin{document}

\title{On the Mechanical Interaction of Light With Homogeneous Liquids}
\author{Michiel de Reus and Neil V. Budko}
\affiliation{Numerical Analysis, Delft Institute of Applied Mathematics, Faculty of Electrical
Engineering, Mathematics and Computer Science,
Delft University of Technology,
Mekelweg 4, 2628 CD Delft, The Netherlands}
\email{n.v.budko@tudelft.nl}

\date{\today}

\begin{abstract}
We investigate one of the consequences of the three competing models describing the mechanical interaction 
of light with a dielectric medium. According to both the Abraham and Minkowski models
the time-averaged force density is zero inside a homogeneous dielectric, whereas the 
induced-current Lorentz force model predicts a non-zero force density. We argue that the latter force, if exists, 
could drive a hydrodynamic flow inside a homogeneous fluid. Our numerical experiments
show that such flows have distinct spatial patterns and may influence the dynamics of particles 
in a water-based single-beam optical trap.
\end{abstract}

\pacs{41.20.Jb, 42.50.Wk, 47.61.Fg}

\maketitle

\section{Introduction}
The mechanical force exerted by light on dielectric objects enables capturing, manipulation,
and analysis of particles at the micro- and nano-meter scales \cite{Ashkin1986,Dienerowitz2008}. 
Often, especially in biological applications, the experimental set-up of an optical trap 
involves a liquid host medium \cite{Molloy2002}. The dynamics of particles in such  
traps is influenced not only by the optical and Brownian forces but by the drag force of
the liquid as well. If operating in the neighborhood of an absorption band, 
a significant thermal flow may be induced by the light itself, and needs to be taken 
into account \cite{Peterman2003, Weinert2008, Schermer2011}. The analysis of the host-medium 
hydrodynamics is important for understanding the observed non-conservative motion of 
trapped particles \cite{Wu2009}, which at the moment is mostly 
attributed to the presence of the so-called scattering force, Brownian `motors' \cite{Khan2011}, 
and thermal/thermophoretic flows. The emergent field of optofluidics also requires a better understanding
of the mechanical interaction of light with liquids \cite{Psaltis2006,Louchev2008,Yang2009,Ryu2010}.

A question arises whether absorption/extinction 
is the only mechanism by which a light-driven hydrodynamic flow may be induced.
Can a flow be created in a lossless liquid?
Since the host liquid is typically optically homogeneous the question can be narrowed down 
to the mechanical forces exerted by the electromagnetic field inside a homogeneous dielectric.
Yet, no simple answer is available at the moment. 
The two dominant approaches, associated with the names of Abraham and Minkowski,
although differing on the form of the mechanical momentum of the electromagnetic field, 
seem to agree that the time-averaged force density (the Helmholtz force density) 
in a homogeneous medium should be exactly zero \cite{Frias2012}. On the contrary, the third approach,
that treats the induced (polarization) currents on equal footing with the externally imposed ones, leads to 
a non-zero Lorentz force density. Thus, one faces a fundamental question: does the electromagnetic
field exert a mechanical force on the polarization currents induced by that same field?

It has been shown that all three approaches give the same
total time-averaged mechanical force on a finite homogeneous solid body, even though the Helmholtz force 
density contributes only at the boundary surface while the Lorentz force density acts across the whole volume of the 
object \cite{Dienerowitz2008}. However, recently it has been suggested that a deformable homogeneous body would respond 
differently to these forces \cite{Frias2012}. Although, the internal stresses induced by light in thin films
may appear to be difficult to detect experimentally,
if such internal stresses do exist, they might induce hydrodynamic flows that are much easier
to observe and should, in fact, be taken into account during the optical trapping.

The quantitative analysis of light-driven flows for a typical single-beam optical trap presented here
is aimed at working out the consequences of the induced-current Lorentz force density model.
We consider both the transparent and absorptive spectral regions.
Our simulations demonstrate that according to this model the light-driven flows 
should be significant (in the order of tens of micrometers per second) even with small 
laser powers ($\sim 1$~mW). Such flows have very distinctive patterns that could be detected
in an experiment.

The paper is organized as follows. First, we briefly describe the three forms of the 
electromagnetic momentum conservation law and the associated time-averaged force densities.
We devote a separate section to the question of modeling of the electromagnetic field in a
single-beam optical trap, where we show that a standard Gaussian beam model is not applicable
for the purposes of force computations as it fails to satisfy the momentum conservation law.
We resort to an alternative method that mimics the Gaussian beam by a linear superposition
of exact fundamental solutions of the Maxwell equations.
Then, we introduce the mechanical coupling between the electromagnetic field and the host liquid
via the body force term of the incompressible Navier-Stokes equations and the Boussinesq approximation 
for the thermally-driven flows. Finally, we present the results of
numerical experiments and our conclusions.

\section{Models of the Electromagnetic Force Density}
There exist two mathematically equivalent views on the electromagnetic field in 
media. In vacuum the Maxwell equations have the form:
\begin{align}
 \label{eq:Maxwell}
\begin{split}
-\nabla\times\bH +\varepsilon_{0}\partial_{t}\bE&=-\bJ^{\rm ext},
\\
\nabla\times\bE +\mu_{0}\partial_{t}\bH&=-\b0,
\end{split}
\end{align}
with $\bE(\bx,t)$, $\bH(\bx,t)$ denoting the 
electric and magnetic field strengths, and $\bJ^{\rm ext}(\bx,t)$ -- 
the external (i.e., independent of the field) electric current density.

A medium different from vacuum can be introduced either via the concept
of electric and magnetic fluxes $\bD$ and $\bB$ entering the Maxwell equations as
\begin{align}
 \label{eq:MaxwellDB}
\begin{split}
-\nabla\times\bH +\partial_{t}\bD&=-\bJ^{\rm ext},
\\
\nabla\times\bE +\partial_{t}\bB&=-\b0,
\end{split}
\end{align}
or via the induced electric and magnetic current densities
$\bJ^{\rm ind}$ and $\bK^{\rm ind}$ as
\begin{align}
 \label{eq:MaxwellInduced}
\begin{split}
-\nabla\times\bH +\varepsilon_{0}\partial_{t}\bE&=-\bJ^{\rm ext}-\bJ^{\rm ind},
\\
\nabla\times\bE +\mu_{0}\partial_{t}\bH&=-\bK^{\rm ind},
\end{split}
\end{align}
Supplied with appropriate constitutive relations, for example, 
\begin{align}
 \label{eq:ConstRelLossless}
 \begin{split}
 \bD&=\varepsilon\bE,\;\;\;\bB=\mu\bH,
\\
 \bJ^{\rm ind}&=(\varepsilon-\varepsilon_{0})\partial_{t}\bE,\;\;\;\bK^{\rm ind}=(\mu-\mu_{0})\partial_{t}\bH,
\end{split}
\end{align}
both formulations lead to exactly the same Maxwell's equations for the fields $\bE$ and $\bH$.

In general, the momentum conservation law has the form:
\begin{align}
 \label{eq:MomentumConservationGeneral}
 \nabla\cdot{\mathbb T}-\partial_{t}\bP = \bff,
\end{align}
where ${\mathbb T}$ is the stress tensor density, $\bP$ is the momentum density, 
and $\bff$ is the force density, which is our main concern. The actual form of this law depends on 
the definition and interpretation of the terms. For example, in the lossless case described 
by the constitutive relations (\ref{eq:ConstRelLossless}) the Abraham and Minkowski expressions for
the force density in the part of the domain without external currents/charges are
\begin{align}
 \label{eq:ForceAbraham}
 \bff^{\rm A}&=\bff^{\rm H}+\frac{\varepsilon\mu-1}{c^{2}}\partial_{t}(\bE\times\bH),
\\
 \label{eq:ForceMinkowski}
 \bff^{\rm M}&=\bff^{\rm H},
\end{align}
where 
\begin{align}
 \label{eq:ForceHelmholtzLossless}
 \bff^{\rm H}=-\frac{1}{2}(\bE\cdot\bE)\nabla\varepsilon-\frac{1}{2}(\bH\cdot\bH)\nabla\mu
\end{align}
is the Helmholtz force density. In a general medium this force density is defined as: 
\begin{align}
 \label{eq:ForceHelmholtz}
 \bff^{\rm H}=\frac{1}{2}\left(\bD\cdot\nabla\bE-\bE\cdot\nabla\bD\right)
+\frac{1}{2}\left(\bB\cdot\nabla\bH-\bH\cdot\nabla\bB\right).
\end{align}
Obviously, after time averaging over a period of harmonic
oscillations the Helmholtz force will be the only non-zero contribution to both 
the Abraham and Minkowski force densities and it will be zero
in a homogeneous lossless medium, i.e.,
\begin{align}
 \label{eq:ForceAMaveraged}
 \langle\bff^{\rm A}\rangle=\langle\bff^{\rm M}\rangle=\langle\bff^{\rm H}\rangle=0,
\;\;\;\varepsilon,\mu=\text{const.}
\end{align}
The Abraham and Minkowski expressions follow from the fluxes-based approach to the 
medium (\ref{eq:MaxwellDB}).
If, on the other hand, one starts with the induced-currents approach (\ref{eq:MaxwellInduced}), 
then the force density turns out to be:
\begin{align}
 \label{eq:ForceLorentz}
\begin{split}
\bff^{\rm L}&=\rho_{\rm e}^{\rm ind}\bE+
\rho_{\rm m}^{\rm ind}\bH
+
\mu_{0}\bJ^{\rm ind}\times\bH-\varepsilon_{0}\bK^{\rm ind}\times\bE,
\\
\rho_{\rm e}^{\rm ind}&=\int_{0}^{t}\nabla\cdot\bJ^{\rm ind}\,dt',
\;\;\;
\rho_{\rm m}^{\rm ind}=\int_{0}^{t}\nabla\cdot\bK^{\rm ind}\,dt'.
\end{split}
\end{align}
In a lossless homogeneous medium this generalized Lorentz force density reduces to:
\begin{align}
 \label{eq:ForceLorentzLossless}
\bff^{\rm L}&=\mu_{0}(\varepsilon-\varepsilon_{0})(\partial_{t}\bE)\times\bH-
\varepsilon_{0}(\mu-\mu_{0})(\partial_{t}\bH)\times\bE.
\end{align}
Assuming time-harmonic fields with the angular frequency $\omega$ and
a linear non-magnetic homogeneous dispersive medium we perform the averaging of the general expression 
(\ref{eq:ForceLorentz}) over a period of oscillations and arrive at the following result:
\begin{align}
 \label{eq:ForceFrequencyDomain}
\begin{split}
\langle\bff^{\rm L}\rangle&=
\frac{1}{2}\omega\mu_{0}(\varepsilon'-\varepsilon_{0})\text{Im}\{\hat{\bS}\}
+\frac{1}{2}\omega\mu_{0}\varepsilon''\text{Re}\{\hat{\bS}\},
\end{split}
\end{align}
where the complex Poynting vector is defined as
\begin{align}
 \label{eq:PoyntingComplex}
\hat{\bS}=\hat{\bE}\times\hat{\bH}^{*},
\end{align}
Here, the complex field amplitudes $\hat{\bE}$, $\hat{\bH}$ satisfy the frequency-domain 
Maxwell's equations, $(\dots)^{*}$ denotes the complex conjugation, and 
$\varepsilon(\omega)=\varepsilon'(\omega)+i\varepsilon''(\omega)$
is the complex permittivity of the host liquid.

The heat produced by the electromagnetic field is given by the following standard 
time-averaged expression derived from the energy conservation law:
\begin{align}
\label{eq:HeatEM}
q=\frac{1}{2}\omega\varepsilon''\hat{\bE}\cdot\hat{\bE}^{*}.
\end{align}

\section{Single-Beam Optical Trap}
\begin{figure*}[t]
\includegraphics[width=17cm]{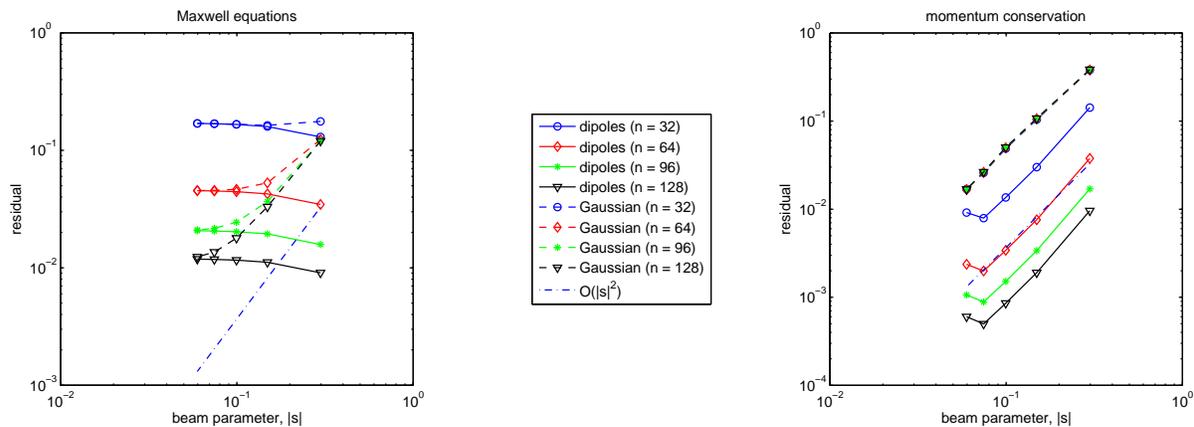}
\caption{Normalized residuals of the Maxwell equations (left) and the momentum conservation 
law (right) as a function of the beam parameter $\vert s\vert$, see eq.~(\ref{eq:GaussianBeamParameter}), 
for different discretization levels ($n=32, 64, 96, 128$ uniformly spaced grid-points along each of the $10\,\mu\text{m}$ 
sides of a cube). 
Solid lines -- dipole-based model, dashed lines -- Gaussian beam approximation. As the beam waist gets wider 
(smaller $\vert s\vert$ values) the residual of the Maxwell equations for the Gaussian 
approximation becomes smaller and is limited by the discretization error. The 
residual of the momentum conservation law stays relatively high irrespectively of the discretization level
for the Gaussian approximation, whereas it is consistently smaller and improves with finer discretization levels 
for the dipole model of the beam.}
\label{fig:Residuals}
\end{figure*}
In optics it is common to model the spatial distribution of the field of a single-beam optical trap 
as a tightly focused Gaussian beam. Unfortunately, this model is not really suitable for the calculation of the 
force density since it violates the conservation of momentum to an extent that may
invalidate the results of numerical experiments. Indeed, the expression
(\ref{eq:ForceFrequencyDomain}) represents the electromagnetic 
force density only if the field quantities appearing in it are the exact solutions of the frequency-domain 
versions of the Maxwell equations (\ref{eq:MaxwellDB}) or (\ref{eq:MaxwellInduced}),
whereas the Gaussian beam is not an exact solution of the Maxwell equations.

Figure~\ref{fig:Residuals} (left) gives the norm of the residual 
of the frequency-domain Maxwell's equations computed numerically via the finite-volume
approximation of the equations on different progressively refined grids. The horizontal axis is the Gaussian 
beam parameter
\begin{align}
 \label{eq:GaussianBeamParameter}
 \vert s\vert = \frac{\lambda_{0}}{2\pi w_{0} \vert n\vert},
\end{align}
where $\lambda_{0}$ is the vacuum wavelength of light, $n$ is the complex refractive index of the medium,
and $w_{0}$ is the beam waist. Thus, the larger is the relative beam waist $w_{0}/\lambda_{0}$, the smaller is
$\vert s\vert$. The residual of the Gaussian approximation (dashed lines) reduces for larger waists, generally following the 
${\mathcal O}(\vert s\vert^{2})$ trend -- the order of the approximation. It also improves with the grid refinement.
This, however, is merely a consequence of the finite-volume approximation of the spatial derivatives in the Maxwell
equations.

Figure~\ref{fig:Residuals} (right) shows the residual of the frequency-domain 
version of the momentum conservation law (\ref{eq:MomentumConservationGeneral}), where the
left-hand side was estimated using the numerical surface integration over the elementary cells.
Here too the Gaussian approximation (dashed lines) improves with the increase in the beam waist. However, refining the discretization 
does not help any more. Hence, we conclude that the error is mainly due to the analytical mismatch in the spatial 
structure of the fields.

A way to improve the single-beam optical trap model is to represent the electromagnetic field as a superposition
of exact analytical solutions of the Maxwell equations due to elementary dipole sources. 
With the help of these fundamental solutions we can model a focused laser beam by distributing an array of dipoles 
over a plane just outside the computational domain and tuning their amplitudes, phases, and polarizations (dipole moments) so that they reproduce
the spatial pattern of the electric field of an ideal focused Gaussian beam passing through that plane.
Basically, this can be viewed as a numerical implementation of the Huygens principle. 

Solid lines in Figure~\ref{fig:Residuals} demonstrate that the residuals
for the dipole-based model of the beam stay roughly constant for various beam waists 
in the Maxwell equations and become smaller for larger beam waists in the momentum conservation law.
Moreover, refining the discretization helps to reduce both residuals making the dipole-based
beam model numerically superior with respect to the Gaussian approximation when calculating the
force densities, especially for tightly focused beams.

\begin{figure}[t]
\hspace*{-0.3cm}
\includegraphics[width=8.5cm]{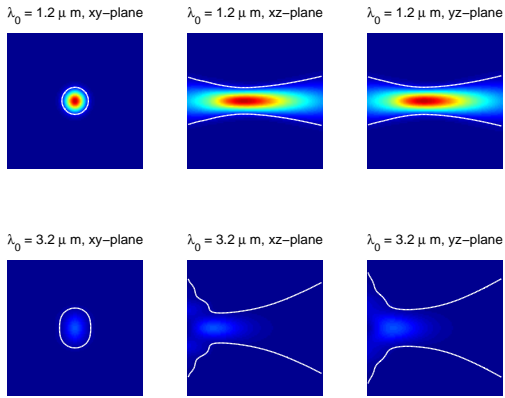}
\caption{Intensity profiles of two focused beams with the same desired waists but different wavelengths 
modeled via the superposition of dipoles tuned to mimic ideal Gaussian beams 
(top: $\lambda_{0}=1.2\,\mu\text{m}$, bottom: $\lambda_{0}=3.2\,\mu\text{m}$).
Actual waists are shown as contour lines.
The larger wavelength beam (bottom row) corresponds to a smaller relative desired waste (i.e. larger 
beam parameter $\vert s\vert$) and deviates from the Gaussian beam while conserving the momentum. 
Both beams have the same power, and the drop in the intensity at $\lambda_{0}=3.2\,\mu\text{m}$ 
is due to the increased absorption in water.}
\label{fig:Beams}
\end{figure}

Our numerical experiments
showed that dipole-based beams very closely resemble ideal Gaussian beams for large waists and contain 
expected diffraction artifacts (side-lobes) with smaller waists. For example, Figure~\ref{fig:Beams} shows
the result for two linearly polarized beams with different wavelengths and the same desired waist. 
The computational domain is a cube with $10\,\mu\text{m}$ sides.
The dipole array is situated $2\,\mu\text{m}$ outside the domain and represents a 
$15\lambda\times15\lambda$ square aperture uniformly filled with $32$ tuned point dipoles in each direction. 
It is important to realize that the dipole model is not always able to achieve the waist size 
of the original Gaussian beam it mimics (see the contour lines in Figure~\ref{fig:Beams}). This is the price one
pays for a better conservation of momentum.

The power emitted by the dipole array into the liquid is determined and adjusted by computing the 
integral of the normal component of the Poynting vector over the (virtual) interface between the computational
domain and the array. Also, in this paper we neglect the reflection of light at the interface between the
vessel and the surrounding medium, considering the whole medium to be optically homogeneous, although the 
liquid is confined to a finite spatial domain. The embarrassingly parallel nature of the superposition principle
allowed us to compute the electromagnetic field and the corresponding force density very efficiently by exploiting
the computational power of the Graphics Processing Unit (GPU). 

\section{Light-Driven Incompressible Fluid}
The Navier-Stokes equations for an incompressible Newtonian fluid are (see e.g. \cite{Batchelor2000})
\begin{align}
\label{eq:DivLaw}
&\nabla\cdot\bv=0,
\\
\label{eq:NavierStokes}
&\rho_{0}\partial_{t}\bv +\rho_{0}(\bv \cdot\nabla) \bv-\mu\nabla^{2} \bv+\nabla p = \bff,
\end{align}
where $\bv(\bx,t)$ is the local velocity of fluid, $p(\bx,t)$ is the pressure, $\rho_{0}$ is the mass density,
$\mu$ is the viscosity and  $\bff(\bx,t)$ is the applied volumetric body force density.
Since light is partially absorbed by the liquid, we need to consider the following heat equation:
\begin{align}
\label{eq:HeatLaw}
&\rho_{0} c_{\rm p}\partial_{t}T+\rho_{0} c_{\rm p}(\bv\cdot\nabla)T
-k\nabla^{2}T=q,
\end{align}
where $T(\bx,t)$ is the temperature, $c_{\rm p}$ is the specific heat capacity at constant pressure,
and $q(\bx,t)$ is the heat source density.

To model the motion of fluid due to the heat-driven expansion we
use the Boussinesq approximation, which amounts to splitting the body force into
two parts:
\begin{align}
\label{eq:Boussinesq}
\bff=\langle \bff^{\rm L}\rangle+\rho \bg,
\end{align}
where $\bg$ is the acceleration due to gravity, and $\langle \bff^{\rm L}\rangle$ is the 
time-averaged Lorentz force density (\ref{eq:ForceFrequencyDomain}).
The modified mass density $\rho(\bx,t)$ is described by
\begin{align}
 \label{eq:DensityLaw}
 \rho=\rho_{0}\left[1-\beta(T-T_{0})\right],
\end{align}
where $T_{0}$ is the reference temperature (before the heat source $q$ is applied), and $\beta$
is the thermal expansion coefficient of the fluid. 

In the present paper we consider the stationary flows only, so that all quantities above are 
considered to be time-independent. This also simplifies the equations (\ref{eq:NavierStokes})
and (\ref{eq:HeatLaw}). We impose the no-slip boundary condition $\bv=0$ at the 
walls of the vessel.

The numerical solution of the above coupled system of non-linear equations was implemented
in the open-source software environment OpenFOAM \cite{OpenFOAM}, which is based on the finite-volume 
discretization scheme and features a rich set of robust algorithms. 
In particular, we employed the iterative SIMPLE (Semi-Implicit Method for Pressure-Linked Equations) algorithm 
\cite{Jasak1996,Patankar1980} that converged reasonably fast to the expected tolerance in all our numerical experiments.
We have also tested the convergence of the numerical solution (to a stable result) as a function of the
discretization step and observed the expected second-order behavior.

\section{Analysis of Flows}
\begin{figure}[t]
\includegraphics[width=8cm]{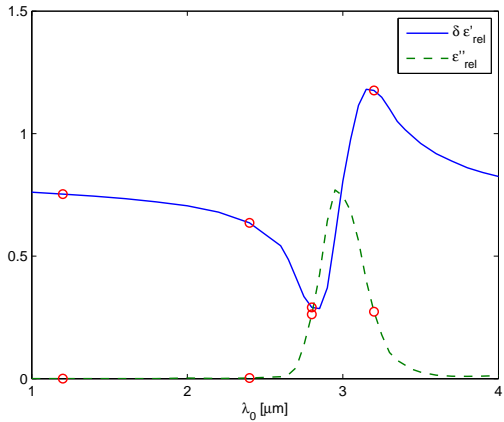}
\caption{Real and imaginary parts of the complex relative permittivity of water in the neighborhood of
the absorption peak at $3\,\mu\text{m}$. Circles indicate different wavelengths ($1.2\,\mu\text{m}$, 
$2.4\,\mu\text{m}$, $2.8\,\mu\text{m}$, and $3.2\,\mu\text{m}$) 
considered in our numerical experiments.}
\label{fig:Dispersion}
\end{figure}
\begin{figure}[t]
\hspace*{-0.5cm}
\includegraphics[width=4cm]{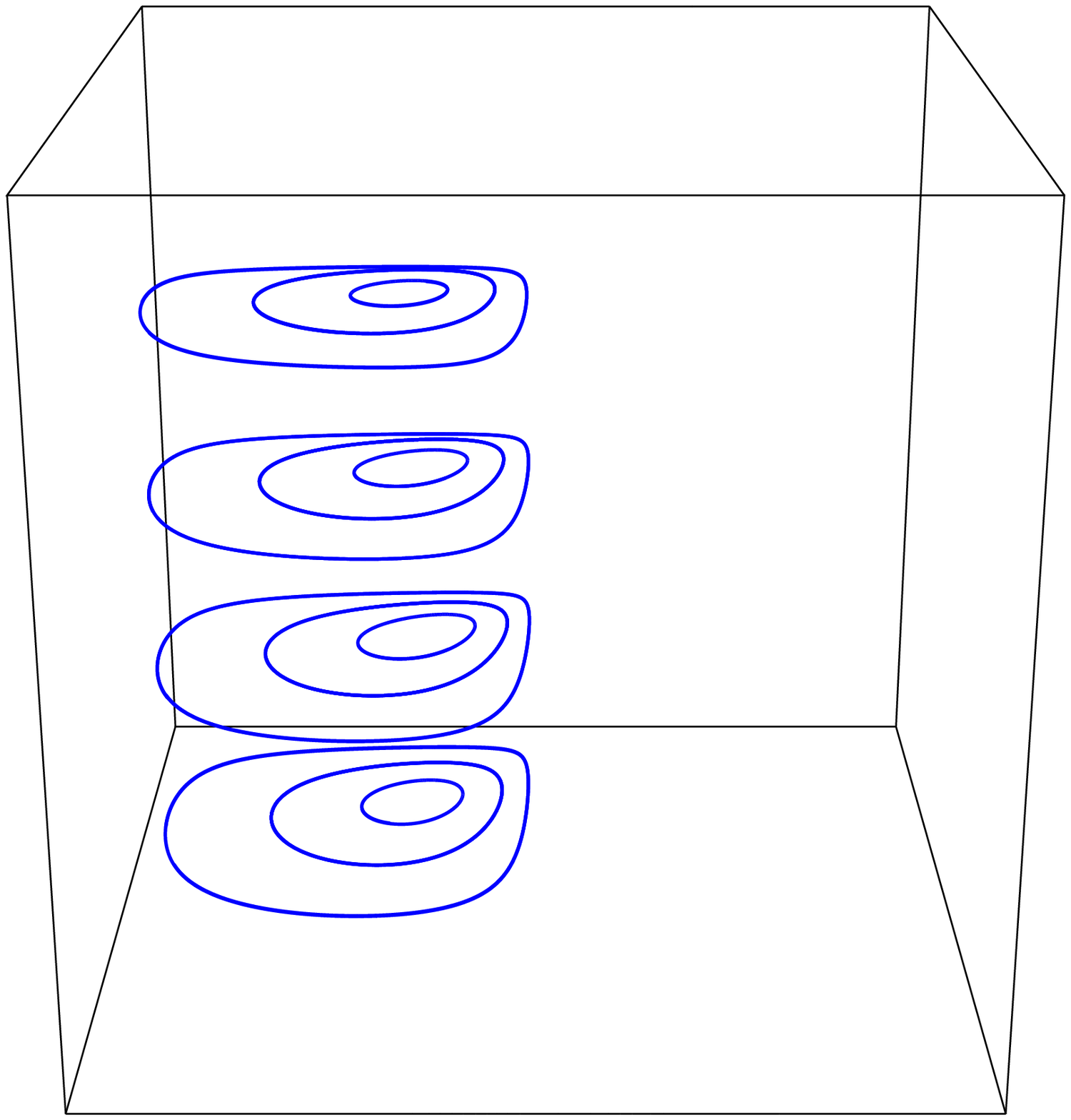}
\includegraphics[width=8cm]{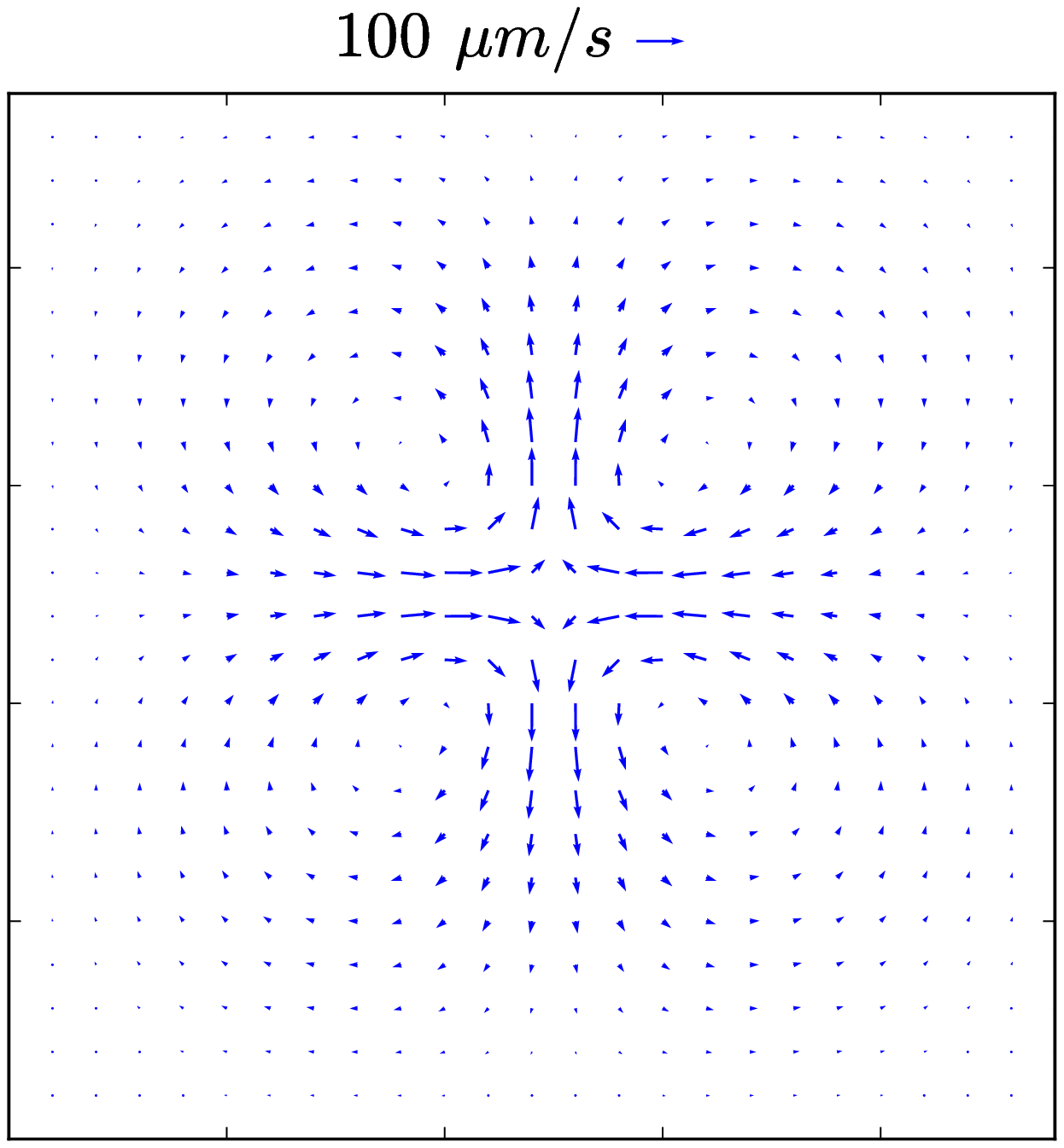}
\caption{Induced flow at $1.2\,\mu\text{m}$ (low optical losses). Top: typical 
stream lines (beam propagates vertically upwards through the middle of the domain). 
Bottom: quiver plot in the plane orthogonal to the beam axis showing the four characteristic quarters.} 
\label{fig:Streams1_2mum}
\end{figure}
\begin{figure*}[t]
\hspace*{-0.5cm}
\includegraphics[width=4.5cm]{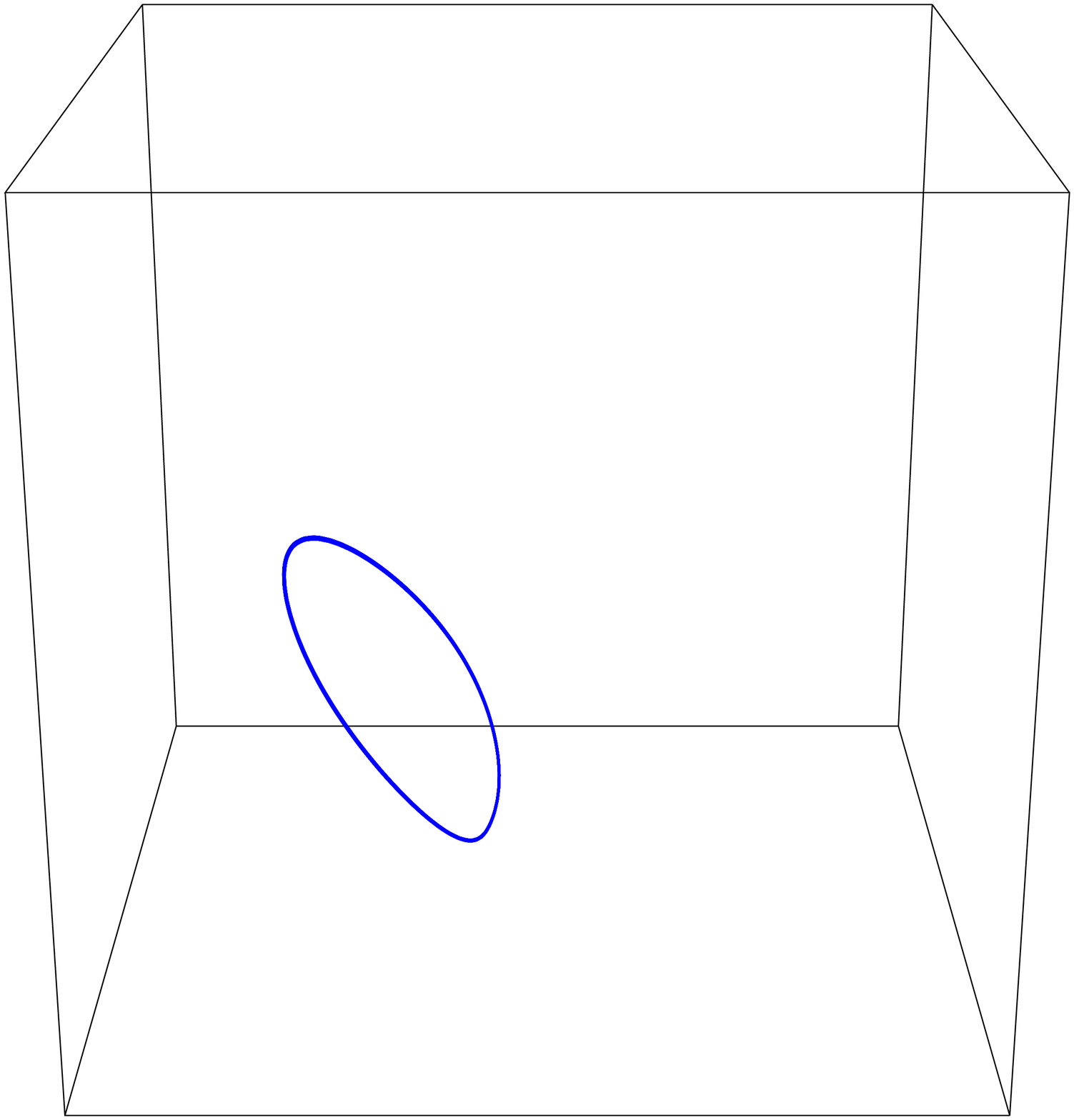}
\includegraphics[width=4.5cm]{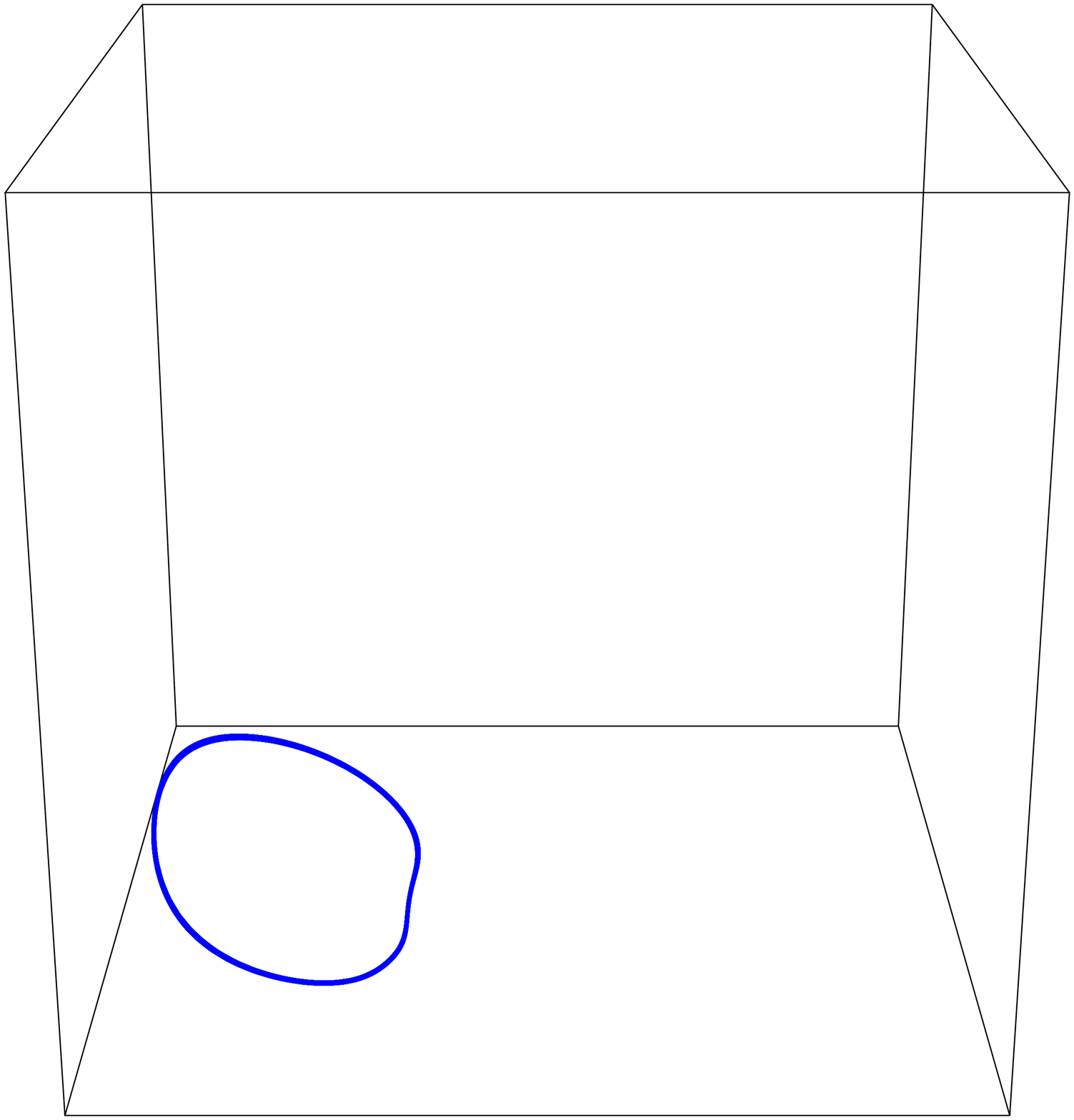}
\includegraphics[width=4.5cm]{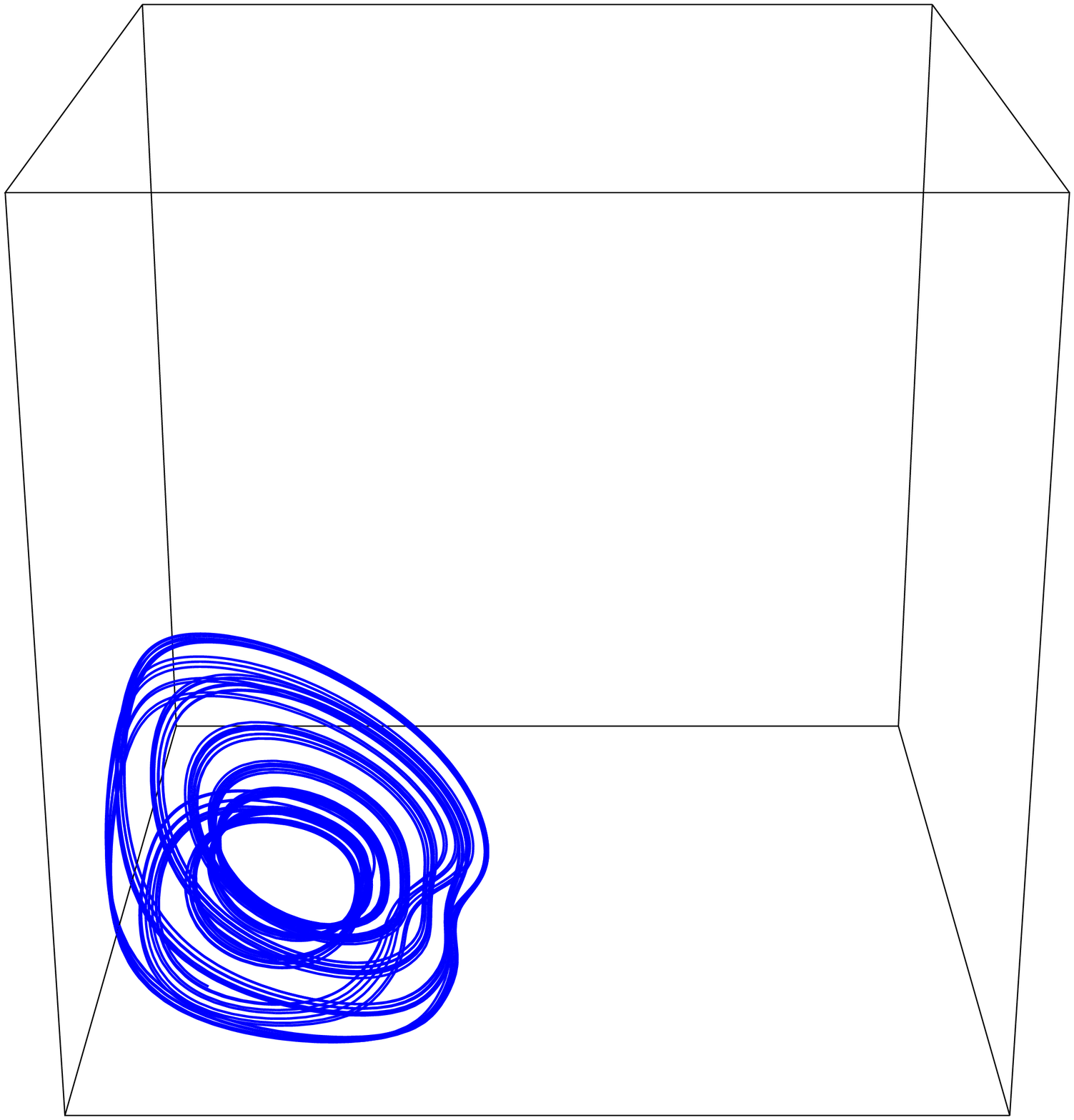}
\includegraphics[width=4.5cm]{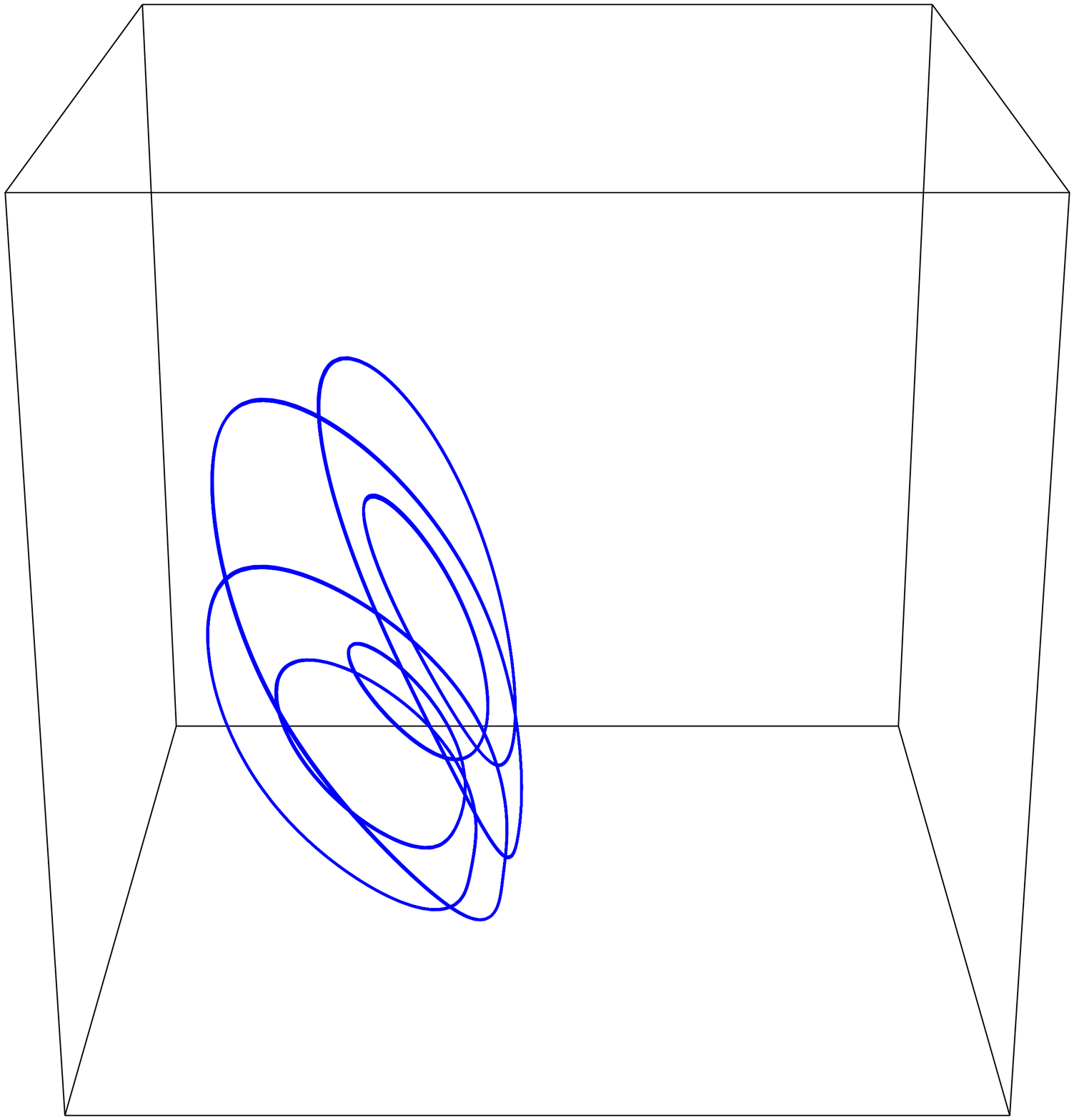}
\hspace*{-0.5cm}
\includegraphics[width=4.5cm]{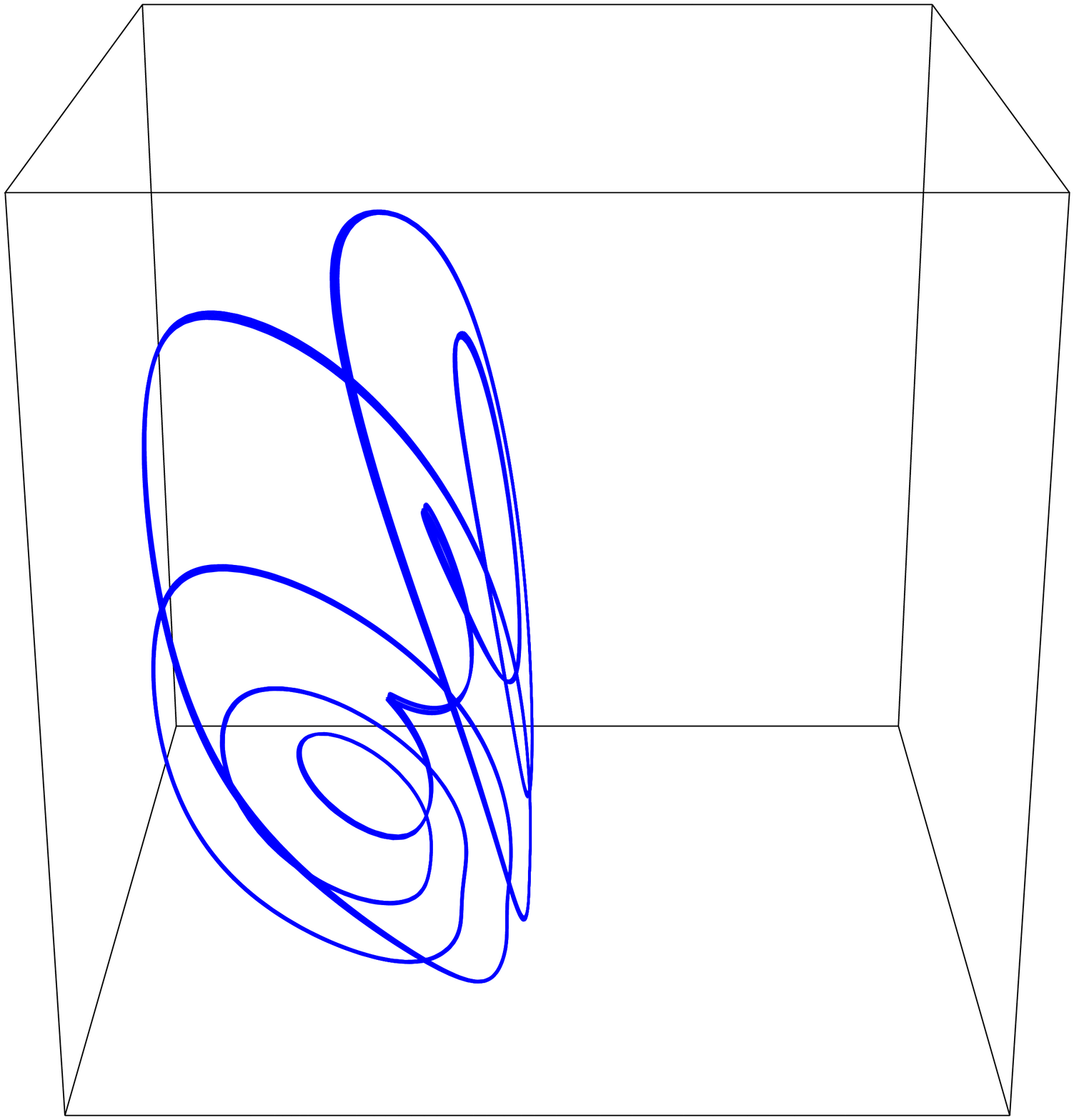}
\includegraphics[width=4.5cm]{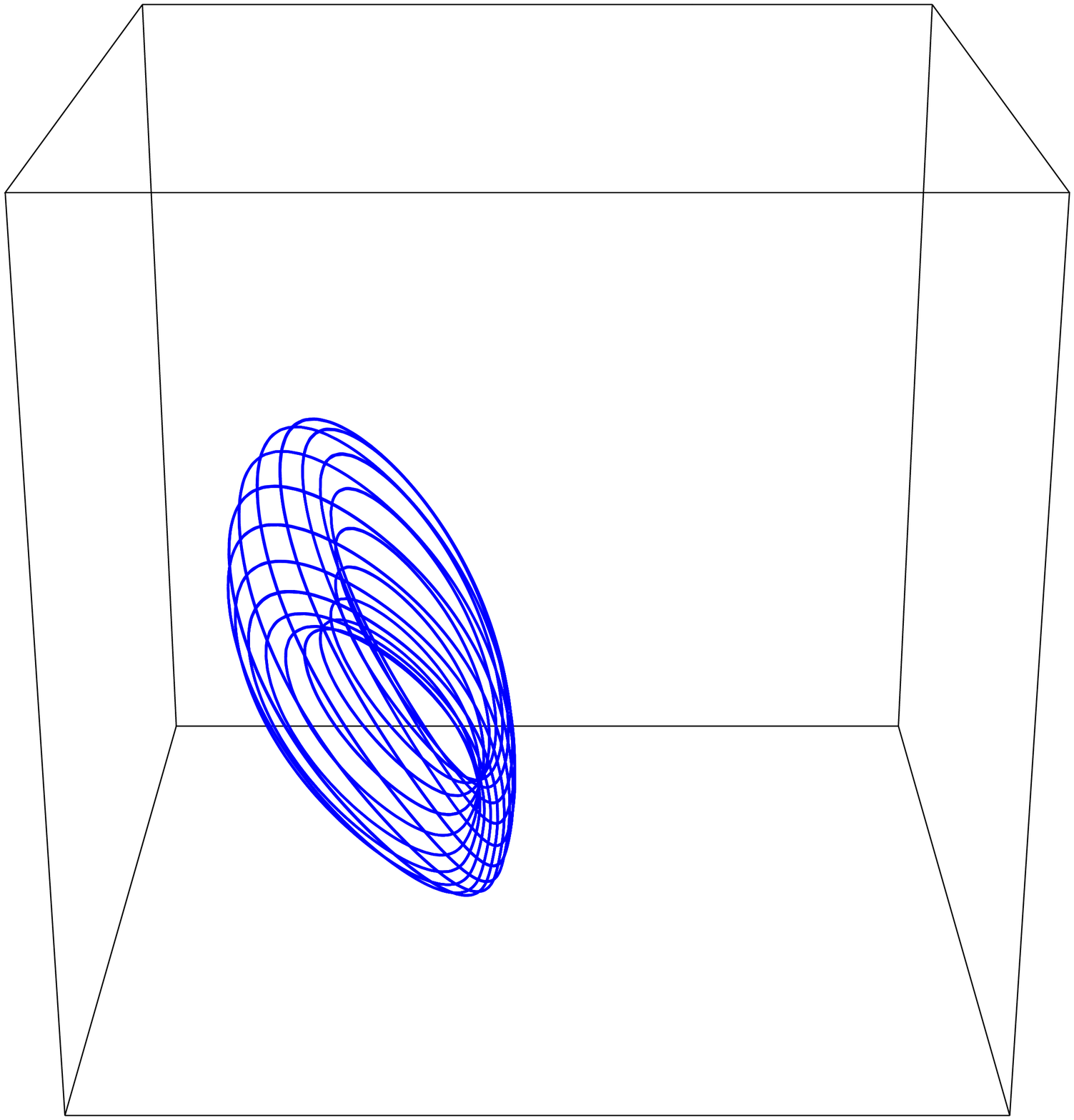}
\includegraphics[width=4.5cm]{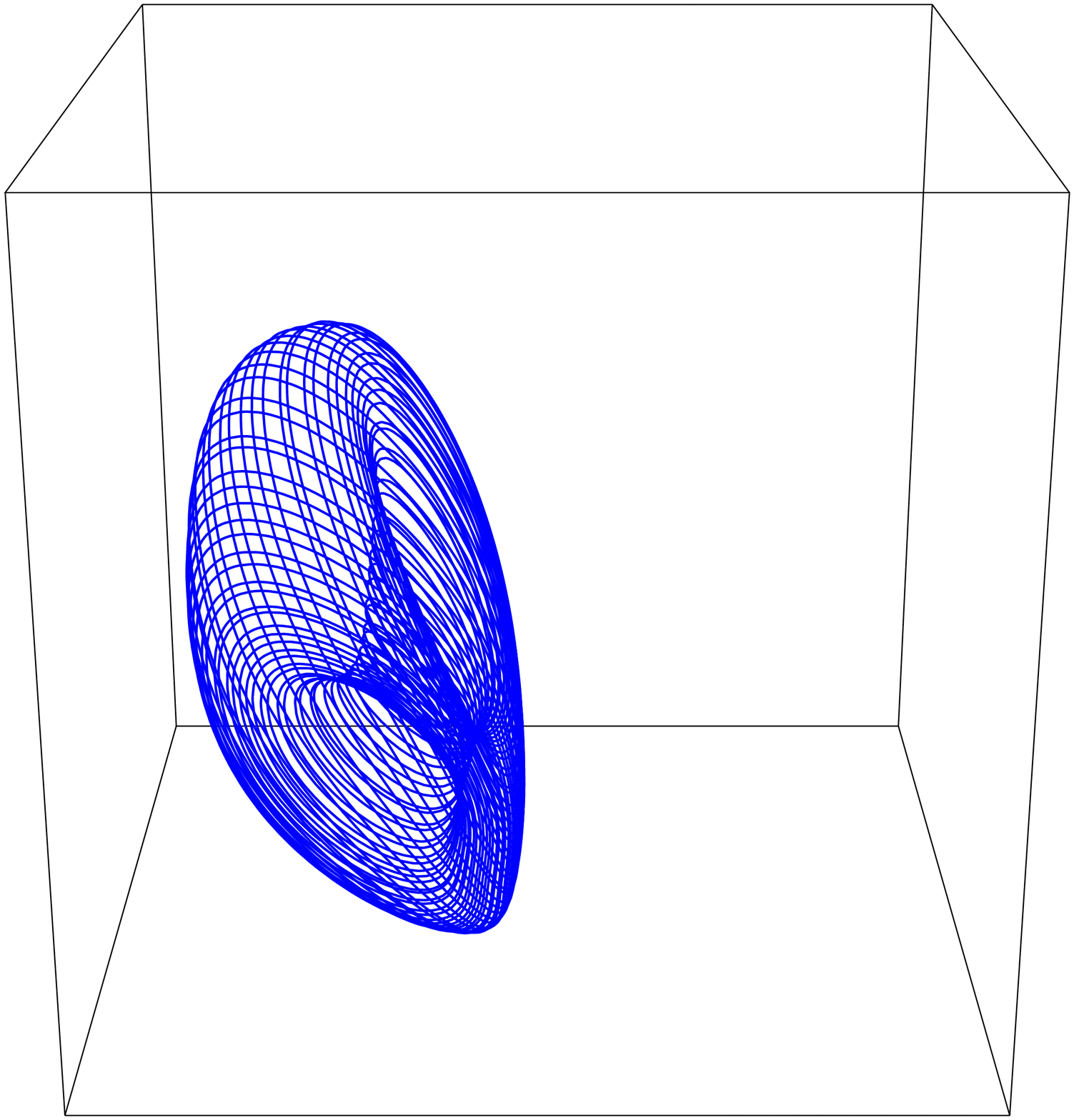}
\includegraphics[width=4.5cm]{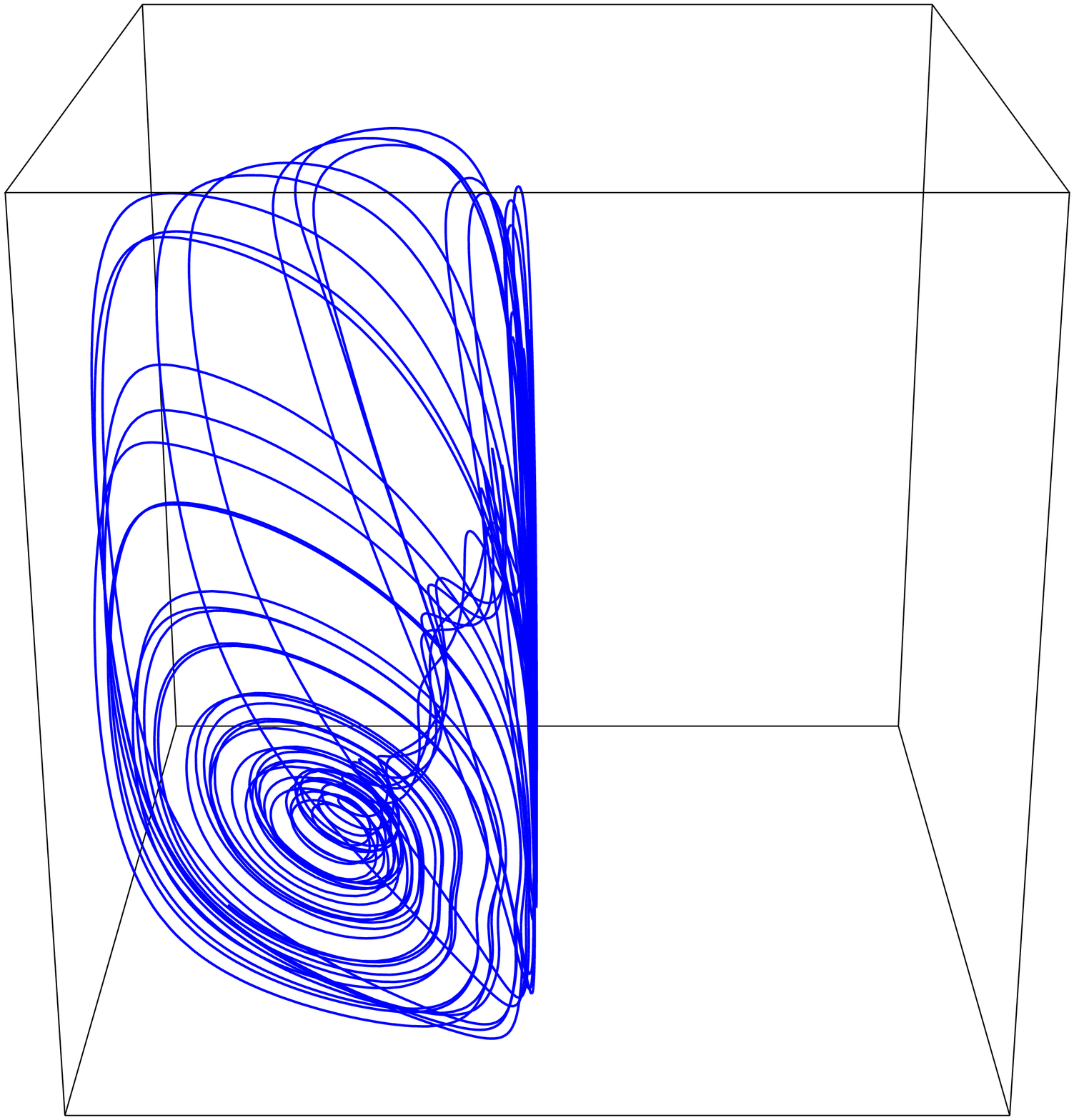}
\caption{Stream lines of the induced flow at $3.2\,\mu\text{m}$ (high optical losses). Each plot shows a single stream line seeded at a particular location and propagated for the same large number of steps with the Runge-Kutta algorithm. In each quarter only two stream lines are simple closed loops (both shown). Other stream lines have a more complex shape (the typical ones are shown). The flow patterns in the other three quarters of the domain are similar. 
}
\label{fig:Streams3_2mum}
\end{figure*}
In this section we describe the flows that would be induced in pure water at the reference temperature $T_{0}=300\,\text{K}$
if the Lorentz force model was indeed true. We have computed the induced flows at several wavelengths 
in the neighborhood of the pronounced absorption peak at $3\, \mu\text{m}$ shown in Figure~\ref{fig:Dispersion}, 
\cite{ref:Hale1973}. The wavelengths $1.2\,\mu\text{m}$, $2.4\,\mu\text{m}$, $2.8\,\mu\text{m}$, and $3.2\,\mu\text{m}$
indicated in Figure~\ref{fig:Dispersion} cover the regions of transparency, as well as normal and anomalous 
dispersion allowing to distinguish between the effects of the different terms in the 
Lorentz force density (\ref{eq:ForceFrequencyDomain}) and the heat source (\ref{eq:HeatEM}).
The most interesting results were obtained at $1.2\,\mu\text{m}$ where
the water is almost lossless and the body force (\ref{eq:ForceFrequencyDomain}) is 
dominated by the first term while the heat source (\ref{eq:HeatEM}) is almost zero, and 
at $3.2\,\mu\text{m}$ where, although the absorption is significant, it does not extinct the beam too soon and 
lets it propagate some length.

The container and the computational domain of our simulations is a 
cube with $10\,\mu\text{m}$ sides. The Gaussian beam (re-modeled by the dipole array) propagates along one of the Cartesian axes coinciding with the edge of the domain 
and is linearly polarized along one of the other edges. We have simulated a whole range of incident powers with the
results presented below corresponding to $1\,\text{mW}$.

In all our experiments the spatial pattern of the flow appears to be divided 
into four quarters by the two planes intersecting at the beam axis.
One of the planes is the polarization plane of the beam and the other plane is orthogonal to it. 
Despite the presence of gravity, which in our case acts along the
polarization direction, i.e., orthogonal to the beam axis, the flow patterns are largely symmetric about 
the mentioned two planes. 

Due to the assumed incompressibility of water the resulting stationary flows are divergence-free. 
Since the water container is closed and there are no sources or sinks the stream lines should be closed. 
This is exactly what we observe in Figure~\ref{fig:Streams1_2mum} where the stream lines and the 
quiver plot of the induced flow at $1.2\,\mu\text{m}$ are shown. The axis of the beam is vertical and 
its direction of propagation is upwards in this and subsequent 3D plots. 
In each of the aforementioned quarters the stream lines represent simple closed 
concentric loops centered around a line parallel to the beam axis. The direction of flow along each such loop is 
always away from the beam axis along the polarization direction and coming back towards the beam along the 
orthogonal direction, see Figure~\ref{fig:Streams1_2mum} (bottom). 
The velocity along the loops is highest at the point closest to the beam axis reaching $\sim 87\,\mu\text{m/s}$.

The flow pattern changes when we consider the wavelengths where the optical absorption in water 
becomes significant introducing the second term in the body force (\ref{eq:ForceFrequencyDomain}). 
For example, in Figure~\ref{fig:Streams3_2mum} we show some of the stream lines of the flow induced 
at $3.2\,\mu\text{m}$. There are just two simple closed loops in each quarter in this case 
(both are shown). The rest of the stream lines have a more complicated shape featuring a large number 
of windings. It is also apparent that the stream lines are no longer orthogonal to the beam axis. 
The direction of flow is determined by both the polarization and the propagation direction of the beam 
in this case. Although the heat term (\ref{eq:HeatEM}) is no longer zero and the direction of gravity breaks
the symmetry of the problem, the overall flow pattern remains almost symmetric about the 
same two orthogonal planes as in the lossless case.

Our simulations show that the flows induced by the beams with higher powers have the same spatial pattern 
as low-power flows and the computed velocity scales linearly with the body force up to $50\,\text{mW}$. 
This is a clear indication of the low Reynolds number regime, meaning that in the future studies a linear Stokes 
approximation can be used to simplify the computational model.

\section{Conclusions}
We have demonstrated that the induced Lorentz force model 
(\ref{eq:ForceLorentz})--(\ref{eq:ForceFrequencyDomain}) results in significant hydrodynamic flows
in the neighborhood of a single-beam optical trap in a liquid environment.
The existence of such flows would suggest the form of the electromagnetic momentum
conservation law (\ref{eq:MomentumConservationGeneral}) different from the ones due to
Abraham and Minkowski, which after time averaging both feature the Helmholtz force density
that equals zero in a homogeneous liquid.
Of course, the fact that such flows have not been directly observed so far (except, perhaps, in 
\cite{Ryu2010,Schermer2011,Khan2011}) could simply indicate their absence.
These induced flows, however, are almost invisible by their nature and may be difficult to detect 
with other means.

Normally a flow pattern can be visualized using small tracer particles, such as ink. This is not straightforward 
in the present case, since both the flow and the tracer particle will be influenced by the laser beam.
The equation of motion for a small spherical particle with mass $m$ and position $\bx(t)$ is
\begin{align}
 \label{eq:NewtonLaw}
 m\frac{d^{2}\bx}{dt^{2}}=\bF_{\rm em} + \bF_{\rm drag},
\end{align}
where $\bF_{\rm em}$ and  $\bF_{\rm drag}$ are the electromagnetic force and
the fluid drag force acting on the particle. 

Considering a sufficiently small tracer particle without optical losses the electromagnetic force can be
approximated by the gradient force:
\begin{align}
 \label{eq:GradientForce}
 \bF_{\rm em}\approx\bF_{\rm grad}=\pi r^{3} {\rm Re}(\varepsilon_{\rm w})
\frac{\varepsilon_{\rm p}/\varepsilon_{0}-1}{\varepsilon_{\rm p}/\varepsilon_{0}+2}\nabla \vert \bE\vert^{2},
\end{align}
where $\varepsilon_{\rm w}$ and $\varepsilon_{\rm p}$ are the permittivities of the water and particle respectively.
The Stokes drag is given by $\bF_{\rm drag}=6\pi\mu_{\rm w} r \bu$,
where $\mu_{\rm w}$ is the dynamic viscosity of water, $r$ is the particle radius, and $\bu$ is the particle velocity 
with respect to the stationary fluid. In the present case, however, the fluid is already in motion due to the
induced flow. Let the velocity of this flow with respect to the stationary reference frame be $\bv$. Then,
the drag force is $\bF_{\rm drag}=6\pi\mu_{\rm w} r (\bv- d\bx/dt)$, and the equation of motion becomes
\begin{align}
 \label{eq:EquationOfMotion}
\begin{split}
m\frac{d^{2}\bx}{dt^{2}}&+6\pi\mu_{\rm w} r\frac{d\bx}{dt}=
\\
&\pi r^{3} {\rm Re}(\varepsilon_{\rm w})
\frac{\varepsilon_{\rm p}/\varepsilon_{0}-1}{\varepsilon_{\rm p}/\varepsilon_{0}+2}\nabla \vert \bE\vert^{2} 
+ 6\pi\mu_{\rm w} r\bv.
\end{split}
\end{align}
Paradoxically, if the tracer velocity $d\bx/dt$ coincides with the velocity $\bv$ of the fluid flow, i.e., the tracer serves its purpose 
and moves with the flow, then the drag force disappears, the tracer will be pulled by the optical gradient force and 
thus no longer follow the flow. On the other hand, if the gradient force pulls the
particle in the direction opposing the flow, the friction term will slow it down and the drag force will 
reduce the trapping efficiency.

Fortunately, the difference in the dependence on the particle radius ($r^{3}$ versus $r$) 
means that sufficiently small tracers will mostly follow the induced flow, since the gradient optical force 
acting on them will be much smaller than the drag force. However, such tracers
will have to be much smaller than the wavelength making it difficult to actually {\it see} their motion.

Finally, the equation of motion (\ref{eq:EquationOfMotion}) does not include the stochastic Brownian term which
will make tracers jump from one streamline to another, further complicating the detection of the simulated 
flow patterns. Hence, a detailed analysis of the particle dynamics and a robust experimental technique 
are needed before the induced flows predicted in this paper can be confirmed or truly ruled out.



\begin{thebibliography}{99}

\bibitem{Ashkin1986}
 A.~Ashkin, J.~M.~Dziedzic, J.~E.~Bjorkholm, and S.~Chu, 
Observation of a single-beam gradient force optical trap for dielectric particles, 
{\em Opt. Lett.}, {\bf 11} 288--290, 1986.

\bibitem{Dienerowitz2008}
M.~Dienerowitz, M.~Mazilu, and K.~Dholakia, 
Optical manipulation of nanoparticles: a review, 
{\em Journal of Nanophotonics}, {\bf 2}, 021875, 2008.

\bibitem{Molloy2002}
J.~E.~Molloy and M.~J.~Padgett, 
Lights, action: optical tweezers, 
{\em Contemporary Physics}, {\bf 43}, 241--258, 2002.

\bibitem{Peterman2003}
E.~J.~G.~Peterman, F.~Gittes, and C.~F.~Schmidt,
Laser-induced heating in optical traps, 
{\em Biophysical Journal}, {\bf 84}, 1308--1316, 2003.

\bibitem{Schermer2011}
R.~T.~Schermer, C.~C.~Olson, J.~P.~Coleman, and F.~Bucholtz,
Laser-induced thermophoresis of individual particles in a viscous liquid,
{\em Optics Express}, {\bf 19}, 10571, 2011.

\bibitem{Weinert2008}
F.~M.~Weinert and D.~Braun, 
Optically driven fluid flow along arbitrary microscale patterns using thermoviscous expansion, 
{\em J. Appl. Phys.}, {\bf 104}, 104701, 2008.

\bibitem{Wu2009}
P.~Wu, R.~Huang, C.~Tischer, A.~Jonas, and E.-L.~Florin,
Direct measurement of the nonconservative force field generated by optical tweezers,
{\em Phys. Rev. Lett.}, {\bf 103}, 108101, 2009.

\bibitem{Khan2011}
M.~Khan and A.~K.~Sood, 
Tunable Brownian vortex at the interface,
{\em Phys. Rev. E}, {\bf 83}, 041408, 2011.

\bibitem{Psaltis2006}
D.~Psaltis, S.~R.~Quake, and C.~Yang, 
Developing optofluidic technology through the fusion of microfluidics and optics, 
{\em Nature}, {\bf 442}, 381--386, 2006.

\bibitem{Louchev2008}
O.~A.~Louchev, S.~Juodkazis, N.~Murazawa, S.~Wada, H.~Misawa,
Coupled laser molecular trapping, cluster assembly, and deposition fed by laser-induced Marangoni convection,
{\em Optics Express}, {\bf 16}, 5673, 2008.

\bibitem{Yang2009}
A.~H.~J.~Yang, T.~Lerdsuchatawanich, and D.~Erickson, 
Forces and transport velocities for a particle in a slot waveguide, 
{\em Nano Lett.}, {\bf  9(3)}, 1182--1188,  2009.

\bibitem{Ryu2010}
J.~C.~Ryu, H.~J.~Park, J.~K.~Park, and K.~H.~Kang, 
New electrohydrodynamic flow caused by the Onsager effect,
{\em Phys. Rev. Lett.}, {\bf 104}, 104502, 2010.

\bibitem{Frias2012}
W.~Frias and A.~I.~Smolyakov,
Electromagnetic forces and internal stresses in dielectric media,
{\em Phys. Rev. E}, {\bf 85}, 046606, 2012.

\bibitem{Batchelor2000}
G.~K.~Batchelor, 
{\em An introduction to fluid dynamics},
Cambridge University Press, Cambridge, 2000.

\bibitem{OpenFOAM}
{\em OpenFOAM 2.1.1 User Guide},
OpenCFD Limited, {\tt http://www.openfoam.org/docs/user/}, 2012.

\bibitem{Jasak1996}
H.~Jasak, 
{\em Error analysis and estimation for the Finite Volume method with applications to fluid flows}, 
PhD Thesis, Imperial College, University of London, 1996.

\bibitem{Patankar1980}
S.~V.~Patankar, 
{\em Numerical heat transfer and fluid flow}, 
Hemisphere Publishing Corporation, 1980.

\bibitem{ref:Hale1973} 
G.~M.~Hale and M.~R.~Querry,
Optical Constants of Water in the 200-nm to 200-$\mu$m Wavelength Region,
{\em Appl. Opt.}, {\bf 12(3)}, 555--563, 1973.

\end{thebibliography}
\end{document}